\documentclass[10pt]{IEEEtran}
\IEEEoverridecommandlockouts
\usepackage{cite}
\usepackage{amsmath,amsthm,amssymb,subfigure,graphicx,multirow}%
\usepackage{algorithmic}
\usepackage{color}
\usepackage{graphicx}
\usepackage{textcomp}
\usepackage{xcolor}
\usepackage{mathtools, cuted}
\usepackage{algorithm,algorithmic}
\usepackage{verbatim,mdframed}
\usepackage{listings,color,soul}

\def\BibTeX{{\rm B\kern-.05em{\sc i\kern-.025em b}\kern-.08em
    T\kern-.1667em\lower.7ex\hbox{E}\kern-.125emX}}

\begin{document}

\title{Cooperative Speed Estimation of an RF Jammer in Wireless Vehicular Networks
}

\author{Dimitrios Kosmanos$^1$, Savvas Chatzisavvas$^1$, Antonios Argyriou$^1$ and Leandros Maglaras$^2$\\
$^1$Department of Electrical and Computer Engineering, University of Thessaly\\
$^2$ Faculty of Computing, Engineering and Media, De Montfort University, Leicester, UK\\
} 

\newcommand{\AAA}[1]{\textcolor{red}{{AA: \rm #1}}}
\newcommand{\DK}[1]{\textcolor{orange}{{DK: \rm #1}}}
\maketitle

\begin{abstract}
In this paper, we are concerned with the problem of estimating the speed of an RF jammer that moves towards a group/platoon of moving wireless communicating nodes. In our system model, the group of nodes receives an information signal from a master node, that they want to decode, while the Radio Frequency (RF) jammer desires to disrupt this communication as it approaches them. For this system model, we propose first a transmission scheme where the master node remains silent for a time period while it transmits in a subsequent slot. Second, we develop a joint data and jamming estimation algorithm that uses Linear Minimum Mean Square Error (LMMSE) estimation. We develop analytical closed-form expressions that characterize the Mean Square Error (MSE) of the data and jamming signal estimates. Third, we propose a cooperative jammer speed estimation algorithm based on the jamming signal estimates at each node of the network. Our numerical and simulation results for different system configurations prove the ability of our overall system to estimate with high accuracy and the RF jamming signals and the speed of the jammer.
\end{abstract}

\begin{IEEEkeywords}
Platoon of Vehicles; LMMSE; MSE; MAE; RF Jamming attack; RF Jammer Speed;
\end{IEEEkeywords}

\section{Introduction}

Wireless communication has constraints in terms of power, bandwidth, reliability, and communication range. As the utility and usefulness of these networks increase every day, more and more malicious competitors appear and target these networks with  different types of security attacks. Radio frequency (RF) jamming is one method that a malicious node can use to disrupt the transmission between the nodes of a wireless network~\cite{punal2014machine},~\cite{malebary2016jamming}. In this type of attack a signal is used to disrupt the communication via the broadcast medium, as most nodes use one single frequency band. In certain application domains where groups of wireless nodes must communicate reliably in broadcast mode, like drone swarms or platoons of autonomous vehicles~\cite{seeda2019} and applications for dynamic charging of electric vehicles through inter-vehicle communication~\cite{electric-vehicles},~\cite{meds}, an RF jammer can have a profound effect in the operation of the system if it can disrupt wireless communication~\cite{mimo-kosmanos},~\cite{karagiannis}. There are methods to defend against a jamming attack such as spread spectrum communication or increase of transmission power, but they typically incur a high cost (power, bandwidth, or complexity). Another way to defend against an RF jamming attack is for the whole group of nodes to move away from the jammer in a flying ad hoc networks (FANETs) environment~\cite{friendly-jamming-uav1},~\cite{friendly-jamming-uav2} or in a platoon that forms a wireless vehicular network~\cite{friendly-jamming-uav1}.
But to do so the group of nodes, especially in a platoon of vehicles, must be able to estimate the behavior of the jammer~\cite{platoons-jamming},~\cite{relative-speed-estimation}. Of particular interest is its speed relative to the platoon since it reveals whether the jammer is approaching or moving away. The focus of this paper is to derive accurate estimates of the speed of the jammer in a group of wireless moving nodes.

Contrary to seeing RF jamming interference as a problem of an individual node, we propose to address it at the group level since the applications of interest fall into this category. Our first contribution is that we propose to use jointly the data from wireless receivers in platoon nodes for the purpose of estimating the jamming signal and eventually the speed of the jammer. To achieve our goal we design a transmission protocol for the platoon and an associated estimation algorithm. With our protocol in the first time slot the master node does not transmit any useful information so we obtain a \emph{clear observation} of just the jamming signal and the receiver noise, while in the second time slot where the information signal is transmitted we observe an additive form the information signal, the jamming signal, and the noise. 
Under this protocol, we use the Linear Minimum Mean Square Error Estimator (LMMSE) to estimate both the information signal $u$ and the jamming signal $z_i$ for every node $i$ in the platoon. Our main result is a closed-form expression of the Mean Square Error (MSE) of the signal $u$ and the jamming signal $z_i$. The second contribution is a new algorithm that combines the jamming signal estimates received at the nodes of the platoon, so as to achieve an accurate estimate of the jammer speed. 
  
The rest of the paper is organised as follows: in Section~\ref{sec:related} we present related work while in Section~\ref{sec:setup} we describe our system model and the assumptions. In Sections~\ref{sec:algorithm},~\ref{sec:speed-estimation}, we present the proposed joint data, jamming signal, and speed estimation algorithms including all the analytical results. In Sections~\ref{sec:numerical},~\ref{sec:simulation} we present numerical and simulation results. Finally in Section~\ref{sec:conclusions} we conclude this paper. 

\section{Related Work}\label{sec:related}

\textbf{Speed Estimation.} Our literature survey indicates that active vehicle safety systems have not benefited sufficiently from the additional information received from a connected vehicle network so as to design more reliable vehicle speed estimation algorithms \cite{speed-estimation1}, \cite{speed-estimation2}, \cite{speed-estimation3}. M. Pirani et al \cite{distributed-speed-estimation1} introduce distributed algorithms for speed estimation where each vehicle can gather information from other vehicles in the network to be used for speed fault detection and reconstruction. This procedure is used as a bank of information for a single vehicle to diagnose and correct a possible fault in its own speed estimation/measurements. The same approach is also considered in \cite{distributed-speed-estimation2} for a platoon of connected vehicles equipped with Cooperative Adaptive Cruise Control (CACC). Without using a distributed system, the authors in ~\cite{relative-speed-estimation} proposed a method for speed estimation between one transmitter and one receiver. However, none of these approaches take into account the possible RF jamming in the area and are not concerned with the speed of the jammer. In contrast, there is considerable work regarding distributed jamming attack detection, but only a few methods exploit distributed jamming signal estimation.

\textbf{Jamming Detection.} Several works cover the problem of distributed jamming detection (but not estimation) in Multiple-Input-Multiple-Output (MIMO) systems. The majority of these works proposed jamming detection methods with a Generalized
Likelihood Ratio Test (GLRT) in MIMO systems ~\cite{jamming-detection-mimo},~\cite{location-velocity-estimation}. 
The authors in ~\cite{siso-jamming-detect} in order to secure the legitimate communication, proposed a jamming detection method in non-coherent Single-Input-Multiple-Output (SIMO) systems, in which channel statistics are not required. It was shown that the probability of detection initially grows with the number of receive antennas but converges quickly, while the channel statistics from the jammer to the receiver always influence the performance. All of these works use additional hardware (e.g. more antennas) on the transmitter and receiver to detect a jamming attack. More recent works like~\cite{karagiannis} proposed methods for jamming detection in Vehicular Networks (VANETs) with Machine Learning (ML) methods like clustering. The authors proposed new algorithms that can differentiate intentional from unintentional jamming as well as extract specific features of the RF jamming signal. In contrast, our proposed method desires to exploit the distributed environment of multiple receivers to effectively estimate the jamming signal and the jammer speed. 

\textbf{Jamming Estimation.} Distributed estimation (DES) is a topic that has been investigated considerably in the literature. However, to the best of our knowledge no works have considered using DES in a setting where a jamming signal and the jammer speed need to be simultaneously estimated. The most closely related work where DES is used for jamming estimation can be found in~\cite{4} where the authors implemented a joint Successive Interference Cancellation (SIC) decoder and LMMSE estimator for an interfering (jamming) signal. Similarly, the authors in~\cite{anti-jamm} investigate the problem of distributed decoding under a white noise jamming attack. However, the aforementioned methods have as prime goal the correct decoding of the valuable data sent by the transmitter under an interference source.


\textbf{Our Work.} In contrast to all the aforementioned works, this paper proposes a two-stage transmission scheme in which the master node remains silent for one slot out of the two time slots. This method is superior in estimating jointly the data and jamming signal using an LMMSE estimation compared to a baseline system. Subsequently, using this jamming signal estimate, the speed of the jammer is also estimated. All the above are accomplished without any extra hardware such as multiple antennas on the transmitter and the receiver.

\section{System Model and Assumptions}
\label{sec:setup}
\begin{figure}[t]
	\includegraphics[width=0.99\linewidth]{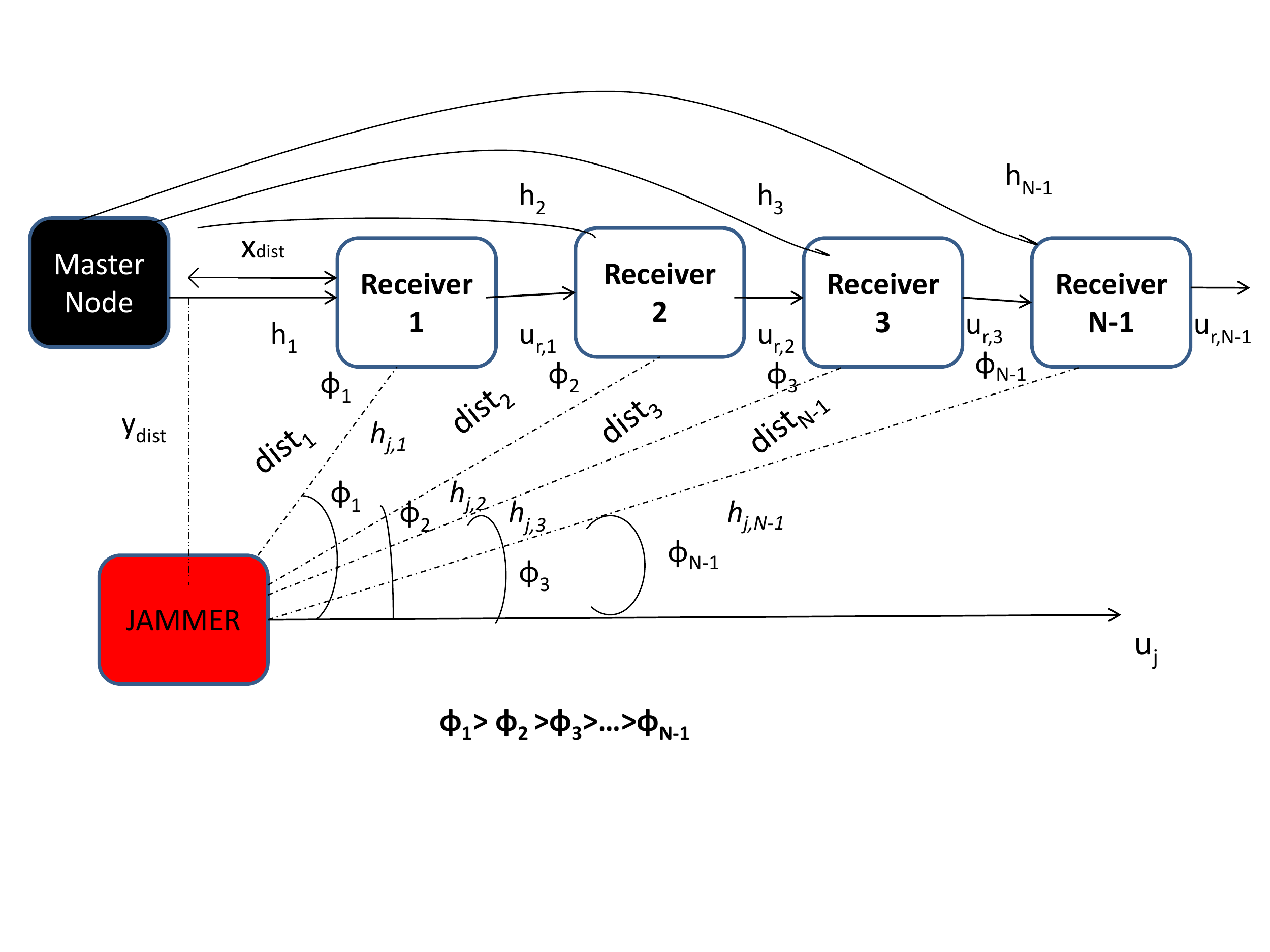}
	\caption{Wireless communication network for DES of the speed of the jammer. }
	\label{fig1}
\end{figure}
\textbf{Topology:} We consider a wireless Vehicle-to-Vehicle (V2V) communication network that consists of a set of $N$ nodes. The first node is the master node who sends the same information messages to the other nodes. Also there is a Jammer (J) who uses his jamming signals to thwart the communication between the master node and the other nodes of the network. 
%
%
%
In our topology the $N -1$ receivers move as a platoon of vehicles with approximately the same speed ($u_{r,1}\simeq u_{r,2}  \simeq ... \simeq u_{r,N-1}=u_r$), using the CACC technology\cite{CACC-2019} and with a constant distance ($d$) between the members of the platoon (fixed to $5m$ in our experiments). Also there is a jammer who moves on a parallel road in relation to the platoon with speed $u_{j}$ and when approaching the platoon within a relatively short distance on the x-axis (at about $x_{dist}$) starts its jamming attack. Observing the topology of the investigated scenario, $N-1$ orthogonal triangles are formed between the jammer, the specific receiver and the vertical projection ($y_{dist}$) of the position of the jammer on the road that the platoon is located. So, for every vertical triangle the Angle of Departure (AOD) values between the jammer and each one from the receivers can be defined using the geometry of the proposed topology as:
 
 \begin{align}
  (1): \cos{\phi_{1}}&=\frac{x_{dist}}{dist_{1}} \nonumber\\
 (2): \cos{\phi_{2}}&=\frac{x_{dist}+d}{dist_{1}} \nonumber\\
 (3): \cos{\phi_{3}}&=\frac{x_{dist}+2*d}{dist_{2}} \nonumber\\
  &... \nonumber\\
 (N-1): \cos{\phi_{N-1}}&=\frac{x_{dist}+(N-2)*d}{dist_{N-1}} \label{eq:geometry1} 
 \end{align}
where $dist_{i}$ is the actual distance between the jammer and the i-th receiver.\par
\textbf{Observation Model:} Each node $i$ during slot $t$ observes $y_{i,t}$ as illustrated in Fig.~\ref{fig1}. In the first time slot, when the master node does not transmit anything, each node observers only the jamming signal $z_i$. In the second time slot when the master node transmits a signal $u$, each node receives two interfering signals: one from the master node, $u$ through a channel $h_i$, and the aggregate signal $z_i$ from the jammer (which is the result of what the jammer transmitted through an unknown channel $h_{j,i}$) that takes into account the relative speed between jammer-receiver and AOD of the transmitted jamming signal. The noise $w_{i,t}$ for each time slot is Additive White Gaussian Noise (AWGN) with zero mean and variance $\sigma_w^2$ and is uncorrelated across the nodes. So in two different time slots we have two observations in every node:
\begin{equation}
y_{i,t}=z_i+w_{i,t} \quad \text{(master\ node\ does not\ transmit)}\label{eqn:1}
\end{equation}
\begin{equation}
y_{i,t}=h_i*u+z_i+w_{i,t} \quad \text{(master\ node\ transmits)}\label{eqn:2}
\end{equation}
In the above $i$ indicates the node and $t$ indicates the time slot. Hence, the observations form the random vector $\vec{y} = [y_{1,1}~y_{1,2}~y_{2,1}~y_{2,2}~...~y_{N-1,1}~y_{N-1,2}]^T$ that has $2(N-1)$ elements. We now define the vectors
\[\vec{u}=[z_1~z_2~z_3~z_4~...~z_{N-1}~u]^T\]
which is a $N\times1$ vector
and
\[\vec{w}=[w_{1,1}~w_{1,2}~w_{2,1}~w_{2,2}~...~w_{N-1,1}~w_{N-1,2}]^T\] which is also a $2(N-1)\times 1$.
The final signal model for our system becomes:
 \begin{equation}
 \label{eq1}
\vec{y}=H \vec{u}+\vec{w}
\end{equation}
where H is the following matrix:

\[
\begin{bmatrix}

1       & 0 & 0 & \dots & 0 \\
1       & 0 & 0 & \dots & h_1 \\
0       & 1 & 0 & \dots & 0 \\
0       & 1 & 0 & \dots & h_2 \\
\hdotsfor{5} \\
0       & 0 & \dots & 1 & 0 \\
0       & 0 & \dots & 1 & h_{N-1}
\end{bmatrix}
\]

 \subsection{Considered Channel Models}
We progressively investigate our idea in the context of more complex channel models and we describe them next.

\textbf{Rayleigh Channel}: For the wireless link we assume flat Rayleigh fading, while the channel remains the same for two consecutive time slots (quasi-static). Hence for every time slot during the transmission of a packet we have $|h_i| \sim Ray(E[|h_i|^2])$~\cite{sno1}. The average received power is $E[|h_i|^2] = 1/dist^{pa}$ where $dist$ is the node's distance from the master node and $pa$ is the path loss exponent set to 3. We assume that the channel  between the master node and the remaining ones is known since it can be easily calculated from packet preambles. 
 

 \textbf{V2V Stochastic Channel}: With this more advanced model, 
 the received signal at the $i \in [1,...,N-1]$ receiver nodes that is received from the jammer through a stochastic wireless V2V channel using the proposed two-slot transmission protocol can be modeled as follows\cite{relative-speed-estimation}:
 \begin{equation}
 \label{V2V_channel:eq1}
y_{i,1}=  \gamma_{i} po_{j,i} e^{j\frac{2\pi}{\lambda} f_{D,i}\tau_{i}}*z_{i} + w_{i,1}    
\end{equation}
\begin{equation}
 \label{V2V_channel:eq2}
y_{i,2}= \gamma_{i} po_{M,i} e^{j\frac{2\pi}{\lambda} f_{DM,i}\tau_{i,M}}*u + \gamma_{i} po_{j,i} e^{j\frac{2\pi}{\lambda} f_{D,i}\tau_{i}}*z_{i} + w_{i,2}    
\end{equation}

All the wireless links between the jammer and the multiple receivers and the links between the master node and the multiple receivers are assumed Line of Sight (LOS). However, the proposed method can be easily applied in a multipath scenario in which in addition to the specular LOS component there are several other Non Line of Sight (NLOS) diffuse components due to multipath reflections~\cite{relative-speed-estimation}.  In the above equations, $\gamma_{i}$ is the amplitude associated with the LOS path, $po_{M,i},po_{j,i}$ represents the corresponding free space propagation losses from the master node and the jammer to the i-th receiver. The $\lambda$ is the wavelength. The complex coefficient $\gamma_{i}$ is assumed to be constant over the observation interval. The variables $\tau_{i,M}, \tau_{i}$ and $f_{DM,i},f_{D,i}$ represent the time delays and Doppler shifts of the transmitted signal from the master node and the jammer, respectively. Finally, $\Delta{u}_{i}$ is the relative speed between the jammer and the specific receiver and $w_{i,1},w_{i,2}$ represents the AWGN with zero mean. Note that~\eqref{V2V_channel:eq1} corresponds to the first time slot in which only the jammer transmits its symbol (as in ~\eqref{eqn:1}) and~\eqref{V2V_channel:eq2} corresponds to the second time slot in which the master node transmits its signal and the jammer interferes too (as in ~\eqref{eqn:2}). The channel model can be modeled exactly as the relation (11) in \cite{relative-speed-estimation}. Since we want to include the relative speed between the jammer and the receiver in the last equations~\eqref{V2V_channel:eq1},~\eqref{V2V_channel:eq2}  we write the Doppler frequency $f_{D,i}$ from the transmitted signal by the jammer as:   
\begin{equation}
\label{doppler}
  f_{D,i}=\frac{\Delta{u}_{i} f_{c}\cos{\phi_{i}}}{c}  
\end{equation}
 where $f_{c}$ the carrier frequency with value $5.9Ghz$ (which is the band dedicated to V2V communication). Also $\cos{\phi_{i}}$ is the incidence AOD between the jammer and the $i$-th receiver and $c$ is the speed of light.
 
 
 
 \subsection{Jammer Behavior}
 We consider jammers that aim to block completely the communication over a link by emitting interference reactively when they detect packets over the air, thus causing a  Denial of Service (DoS) attack. The jammers minimize their activity to only a few symbols per packet and use minimal, but sufficient power, to remain undetected. We assume that the jammer is pretty capable and is able to sniff any symbol of the over the air transmissions in real-time and react with a jamming signal that flips selected symbols at the receiver with high probability (see \cite{detection-reactive}).  This type of reactive jammer is designed to start transmitting upon sensing energy above a certain threshold in order for a reactive jamming attack to succeed. We set the latter to $-75$ dBm as it is empirically determined to be a good tradeoff between jammer sensitivity and false transmission detection rate, when an ongoing 802.11p transmission is assumed ~\cite{punal2014vanets}.
 
For the jamming signal we don't have any information for its variance. We assume that the reactive jammer transmits after its being triggered for two consecutive time slots and this has the result that the jamming signal is the same. Also we assume that the channel conditions between the jammer and the multiple receivers remain the same through two consecutive time slots.

\section{Joint Data and Jamming Signal Estimation}
\label{sec:algorithm}
For estimating the information and the jamming signal in this paper we adopt the LMMSE~\cite{4},~\cite{sno1}. An LMMSE estimator is an estimation method which minimizes the MSE which is a common measure of estimator quality. The  LMMSE estimator ensures the minimum MSE from all linear estimators. For our general linear model $\vec{y}=H\vec{u}+\vec{w}$, the estimator of $\vec{u}$ is given as:
	\begin{equation}
	\label{eq:estimator1}
	\hat{\vec{u}}=(H^HC_w^{-1}H+C_u^{-1})^{-1}H^HC_w^{-1}\vec{y}
	\end{equation}
	where $C_{\vec{w}}$ and $C_{\vec{u}}$ are the auto-covariance matrices of $\vec{w}$ and $\vec{u}$ respectively. The MSE of this estimator is the trace of $C_{\vec{e}}$, that is the covariance matrix or the estimation error:
	\begin{equation}
	MSE=\text{Tr}(C_{\vec{e}})=\text{Tr}((H^HC_w^{-1}H+C_u^{-1})^{-1}))
	\end{equation}
	
\subsection{MSE derivation}
As the literature has shown, a very challenging task is to produce a closed-form expression for the desired estimator and signal model~\cite{4,review1,10}. In this subsection we outline the process that has led to the desired expression that will help us study the behavior of the proposed system.
	
Recall that in our model we assume that the noise is AWGN with zero mean and variance $\sigma_w^2$ and is uncorrelated across the nodes. We have no information about the jamming signal and so we assume that its mean is zero. Under these assumptions and with the use of the general LMMSE estimator, the MSE for the information $u$ and jamming signal $z_i$ for nodes is given in~\eqref{eqn:MSEu}, and~\eqref{eqn:MSEz} respectively.
\begin{figure*}[htb]
	\begin{equation}
	\label{eqn:MSEu}
	MSE_u=\frac{1}{\sum_{n=2}^{N}(\frac{h_n^2}{\sigma_{wn,2}^2})+\frac{1}{s_u^2}-\sum_{n=2}^{N}(\frac{h_n^2}{\sigma_{wn,2}^4*(\frac{1}{\sigma_{wn,1}^2}+\frac{1}{\sigma_{wn,2}^2}+\frac{1}{\sigma_{zn}^2})})}
	\end{equation}
\end{figure*}

\begin{figure*}[!t]
	\begin{equation}
	\label{eqn:MSEz}
	MSE_{zi}=\frac{\prod_{n=2,n\neq{i}}^{N}(\frac{1}{\sigma_{wn,1}^2}+\frac{1}{\sigma_{wn,2}^2}+\frac{1}{\sigma_{zn}^2})*(\sum_{k=2}^{N}(\frac{h_n^2}{\sigma_{wn,2}^2})+\frac{1}{s_u^2})-\sum_{k=2}^{N}(\frac{h_n^2}{\sigma_{wn,2}^4})*\prod_{n=2,n\neq{i,k}}^{N}(\frac{1}{\sigma_{wn,1}^2}+\frac{1}{\sigma_{wn,2}^2}+\frac{1}{\sigma_{zn}^2})}{\prod_{n=2}^{N}(\frac{1}{\sigma_{wn,1}^2}+\frac{1}{\sigma_{wn,2}^2}+\frac{1}{\sigma_{zn}^2})*(\sum_{n=2}^{N}(\frac{h_n^2}{\sigma_{wn,2}^2})+\frac{1}{s_u^2})-\sum_{k=2}^{N}(\frac{h_n^2}{\sigma_{wn,2}^4})*\prod_{n=2,n\neq{i}}^{N}(\frac{1}{\sigma_{wn,1}^2}+\frac{1}{\sigma_{wn,2}^2}+\frac{1}{\sigma_{zn}^2})}
	\end{equation}
\end{figure*}
In order to understand better the implications of the produced expression we present results for the case of $N=4$ where we have that
	\begin{equation}
	MSE_u=\frac{1}{s-\frac{h_2^2}{\sigma_{w2,2}^4*\alpha}-\frac{h_3^2}{\sigma_{w3,2}^4*\beta}-\frac{h_4^2}{\sigma_{w4,2}^4*\gamma}}\label{eq}
	\end{equation}
Also $MSE_{z2}$ is equal to
	\begin{equation}
\frac{\beta*\gamma*s-\frac{h_3^2}{\sigma_{w3,2}^4}*\gamma-\frac{h_4^2}{\sigma_{w4,2}^4}*\beta}{\alpha*\beta*\gamma*s-\frac{h_2^2}{\sigma_{w2,2}^4}*\beta*\gamma-\frac{h_3^2}{\sigma_{w3,2}^4}*\alpha*\gamma-\frac{h_4^2}{\sigma_{w4,2}^4}*\alpha*\beta}
	\end{equation}
	where:
	\[
	s=\frac{h_2^2}{\sigma_{w2,2}^2}+\frac{h_3^2}{\sigma_{w3,2}^2}+\frac{h_4^2}{\sigma_{w4,2}^2}+\frac{1}{s_u^2}
	\]
	\[
	\alpha=\frac{1}{\sigma_{w2,1}^2}+\frac{1}{\sigma_{w2,2}^2}+\frac{1}{\sigma_{z2}^2}
	\]
	\[
	\beta=\frac{1}{\sigma_{w3,1}^2}+\frac{1}{\sigma_{w3,2}^2}+\frac{1}{\sigma_{z3}^2}
	\]
	\[
	\gamma=\frac{1}{\sigma_{w4,1}^2}+\frac{1}{\sigma_{w4,2}^2}+\frac{1}{\sigma_{z4}^2}
	\]
The first thing we notice from these expressions is that the MSE of the information signal $u$ is inversely proportional to the number of nodes, that is we have benefits in the accuracy of bit detection (MSE can be easily converted to Signal-to-Noise Ratio (SNR) and Bit Error Rate (BER)) when more nodes assist in the estimation process. Regarding the MSE of the estimated jamming signal it is also increased with a higher number of nodes but this is not obvious from the expression~\eqref{eqn:MSEz} that is more involved. The precise quantification of these gains is presented in the respective performance evaluation section where we will delve into the performance of this estimator in isolation first.

\section{Jammer Speed Estimation}
\label{sec:speed-estimation}
Our ultimate goal is to estimate the jammer speed based on jamming signal estimates that we obtained from the previous section. Fig~\ref{fig1} indicates that between the jammer and each receiver there is different AOD and a different distance ($dist_{i}$). Using again the estimator in~\eqref{eq:estimator1} we propose to combine the values in the $N\times 1$ estimated vector $\vec{u}$ (that contains the joint data and the jamming signal) by diving them pairwise and taking then the absolute value:
\begin{align*}
\label{eq3}
 |\frac{\hat{u}_{1}}{\hat{u}_{2}}| &= |\frac{\hat{z}_{1}}{\hat{z}_{2}}|=|\frac{h_{j,1} z} {h_{j,2} z}|=|\frac{\gamma_{1} po_{j,1}  e^{j\frac{2\pi}{\lambda} f_{D,1}\tau_{1}}} {\gamma_{2} po_{j,2} e^{j\frac{2\pi}{\lambda} f_{D,2}\tau_{2}}   }| \\
&...\\
|\frac{\hat{u}_{N-2}}{\hat{u}_{N-1}}|&=|\frac{\hat{z}_{N-2}}{\hat{z}_{N-1}}|=|\frac{\gamma_{N-2} po_{j,N-2}  e^{j\frac{2\pi}{\lambda} f_{D,N-2}\tau_{N-2}}} {\gamma_{N-1} po_{j,N-1}  e^{j\frac{2\pi}{\lambda} f_{D,N-1}\tau_{N-1}}   }|
\end{align*}
Recall that 
the $N-1$ receivers are assumed to be close to each other, resulting in a constant value for the free space propagation loss $po_{j,i}$ and the random variable $\gamma_{i}$ for the observation interval in the above equations. Under these assumptions the previous set of equations is simplified to:
\begin{align*}
  \frac{\hat{z}_{1}}{\hat{z}_{2}}&=\frac{e^{j\frac{2\pi}{\lambda} \Delta{u}_{1} \frac{f_c}{c} \cos{\phi_{1}} \tau_{1}}} {e^{j\frac{2\pi}{\lambda} \Delta{u}_{2} \frac{f_c}{c} \cos{\phi_{2}} \tau_{2}}   } \\
&...\\
\frac{\hat{z}_{N-2}}{\hat{z}_{N-1}}&=\frac{e^{j\frac{2\pi}{\lambda} \Delta{u}_{N-2} \frac{f_c}{c} \cos{\phi_{N-2}} \tau_{N-2}}} {e^{j\frac{2\pi}{\lambda} \Delta{u}_{N-1} \frac{f_c}{c} \cos{\phi_{N-1}} \tau_{N-1}}   }
\end{align*}
By taking the natural logarithm of the expressions on the left and right we have:
\begin{align*}
  \ln{(\frac{\hat{z}_{1}}{\hat{z}_{2}})}&=\ln{(\frac{e^{\omega |u_{r,1}-u_{j}| \cos{\phi_{1}}\tau_{1}}}{ e^{\omega |u_{r,2}-u_{j}| \cos{\phi_{2}}\tau_{2}}} )  } \\
& ...\\
 \ln{(\frac{\hat{z}_{N-2}}{\hat{z}_{N-1}})}&=\ln{(\frac{e^{\omega |u_{r,N-2}-u_{j}| \cos{\phi_{N-2}}\tau_{N-2}}}{ e^{\omega |u_{r,N-1}-u_{j}| \cos{\phi_{N-1}}\tau_{N-1}}} )  } 
\end{align*}
where $u_{r,i}$ is the speed of every receiver, $u_{j}$ the speed of the jammer in the area and the variables $f_{cx}=\frac{f_c}{c}$, $\omega=j\frac{2\pi}{\lambda} f_{cx}$. If we assume that the jammer approaches the $i$-th receiver at a speed lower than its own speed the relative speed between jammer and receiver is positive and so $|u_{r,i}-u_{j}|= u_{r,i}-u_{j}$. By simplifying the previous logarithmic equations we have: 
\begin{align*}
\small
  \ln{(\frac{\hat{z}_{1}}{\hat{z}_{2}})}&=\omega [(u_{r,1}-u_{j}) \cos{\phi_{1}}\tau_{1} -(u_{r,2}-u_{j}) \cos{\phi_{2}}\tau_{2}]\\
  &...\\
  \ln{(\frac{\hat{z}_{N-2}}{\hat{z}_{N-1}})}& =\omega [(u_{r,N-2}-u_{j}) \cos{\phi_{N-2}}\tau_{N-2}\\
  &-(u_{r,N-1}-u_{j}) \cos{\phi_{N-1}}\tau_{N-1}]
\end{align*}
In the above equations the estimated jamming signal values on the left-hand side are complex numbers of the form: $\hat{a}_{1}+\hat{b}_{1}j,...,\hat{a}_{N-2}+\hat{b}_{N-2}j$. We observe that the real part of the above equations on the right side is equal to zero. So all the real parts, that is the $\hat{a}$'s, are equal to zero. We also assume that the receivers move at similar speeds ($u_{r,1}\simeq u_{r,2}  \simeq ... \simeq u_{r,N-1}=u_r$) as they are members of the platoon. By replacing the AOD values with the order of equations~\eqref{eq:geometry1} and the time delays as $\tau_{1}=\frac{dist_{1}}{c}, \tau_{2}=\frac{dist_{2}}{c}, ..., \tau_{N-1}=\frac{dist_{N-1}}{c} $ we have:

\begin{align*}
\tiny
 \hat{b}_{1}&=\omega [(u_{r}-u_{j}) \frac{x_{dist}}{dist_{1}}*\frac{dist_{1}}{c} \\
  &-(u_{r}-u_{j}) \frac{x_{dist}+d}{dist_{2}}*\frac{dist_{2}}{c}]\\
  &...\\
  \tiny
  \hat{b}_{N-2}& =\omega [(u_{r}-u_{j}) \frac{x_{dist}+(N-3)*d}{dist_{N-2}}*\frac{dist_{N-2}}{c}\\
  &-(u_{r}-u_{j}) \frac{x_{dist}+(N-2)*d}{dist_{N-1}}*\frac{dist_{N-1}}{c}]
\end{align*}

Now, we can solve the above equations for the speed of the jammer:


\begin{align*}
(1):\hat{u}_{j}&=\frac{\hat{b}_{1}*\lambda*c^{2}}{d*2\pi*f_{c}}+u_{r}\\
(2):\hat{u}_{j}&=\frac{\hat{b}_{2}*\lambda*c^{2}}{d*2\pi*f_{c}}+u_{r}\\
&...\\
(N-2):\hat{u}_{j}&=\frac{\hat{b}_{N-2}*\lambda*c^{2}}{d*2\pi*f_{c}}+u_{r}
\end{align*}
This means that we have $N-2$ equations that involve the speed of the jammer and the known  value  of $d$ which is the distance between the members of the platoon. We observe that the only factor that differentiates these equations are the $(\hat{b}_{1}, \hat{b}_{2} ,..., \hat{b}_{N-2})$ which are the imaginary parts of the estimated complex numbers $(\ln{(\frac{\hat{z}_{1}}{\hat{z}_{2}})}, \ln{(\frac{\hat{z}_{2}}{\hat{z}_{2}})}, ..., \ln{(\frac{\hat{z}_{N-2}}{\hat{z}_{N-1}})})$. Consequently, these values are only related to the estimated jamming signals $\hat{z}_{i}$.  Obtaining the unbiased  sample mean estimator of the above point estimates for the speed of the jammer we have: 
\begin{equation}
\label{sum1}
\hat{\bar{u}}_{j}=\sum_{l=1}^{N-2}{\frac{1}{N-2}(\frac{\hat{b}_{l}*\lambda*c^{2}}{d*2\pi*f_{c}}+u_{r})}
\end{equation}
Hence, if we increase the number of receivers we obtain a better estimate of the speed of the jammer.
Now in the case that the jammer approaches the $i$-th receiver at a speed lower than the relative speed between the jammer and the receiver, it has positive sign if $|u_{r}-u_{j}|= u_{j}-u_{r}$. Repeating the above procedure results in something analogous to~\eqref{sum1}, namely:
\begin{equation}
\label{sum2}
\hat{\bar{u}}_{j}=\sum_{l=1}^{N-2}{\frac{1}{N-2}(u_{r}-\frac{\hat{b}_{l}*\lambda*c^{2}}{d*2\pi*f_{c}})}
\end{equation}
We must note that we do not need to know a-priori the correct sign of the relationship $|u_{r}-u_{j}|$ since one of the two~\eqref{sum1},~\eqref{sum2} will have a negative sign and consequently this estimated jammer speed value must be rejected. In this case, the alternative equation must be used to estimate the speed of the jammer.

\section{Numerical and Simulation Results for AWGN and Rayleigh Channels}
\label{sec:numerical}
For our simulations we assume that the master node together with the other nodes form a platoon of vehicles that move together in a specific direction with approximately a constant velocity. The jammer is in a specific distance and moves in parallel with them but we do not have any information for its position and channel condition between itself and the nodes in the platoon. We gradually present results for the AWGN channel, a Rayleigh fading channel, and finally a realistic vehicular channel that includes LOS and shadowing from obstacles in the next section. In this way we can offer a full exploration of all the aspects of our system.

For the AWGN and Rayleigh channels our purpose is to evaluate the ability of the estimator in~\eqref{eqn:MSEu} and ~\eqref{eqn:MSEz} to accurately estimate the jamming signal. Consequently, we also test a baseline system where the master node transmits data continuously without stopping its transmission as with the proposed scheme. In our analytical model this result can be obtained by setting the noise variance to infinity in~\eqref{eqn:1}. Furthermore, we assume $\sigma^2_{w}$ to be equal to $0.1$. The information signal $u$ is a random binary sequence with power equal to $\sigma_u^2=1$ leading thus to a transmit SNR of 10dB. Higher SNRs would lead to higher gains. For the jamming signal note that its variance $\sigma_{zi}^2$ at every node takes different values because of channel fading. We implemented our algorithm in Matlab and we executed $50000$ iterations for every different system configuration. For our results we present the MSE for the transmitted information $u$ and for the jamming signal $z_i$.

\begin{figure}[t]
\includegraphics[width=0.99\linewidth]{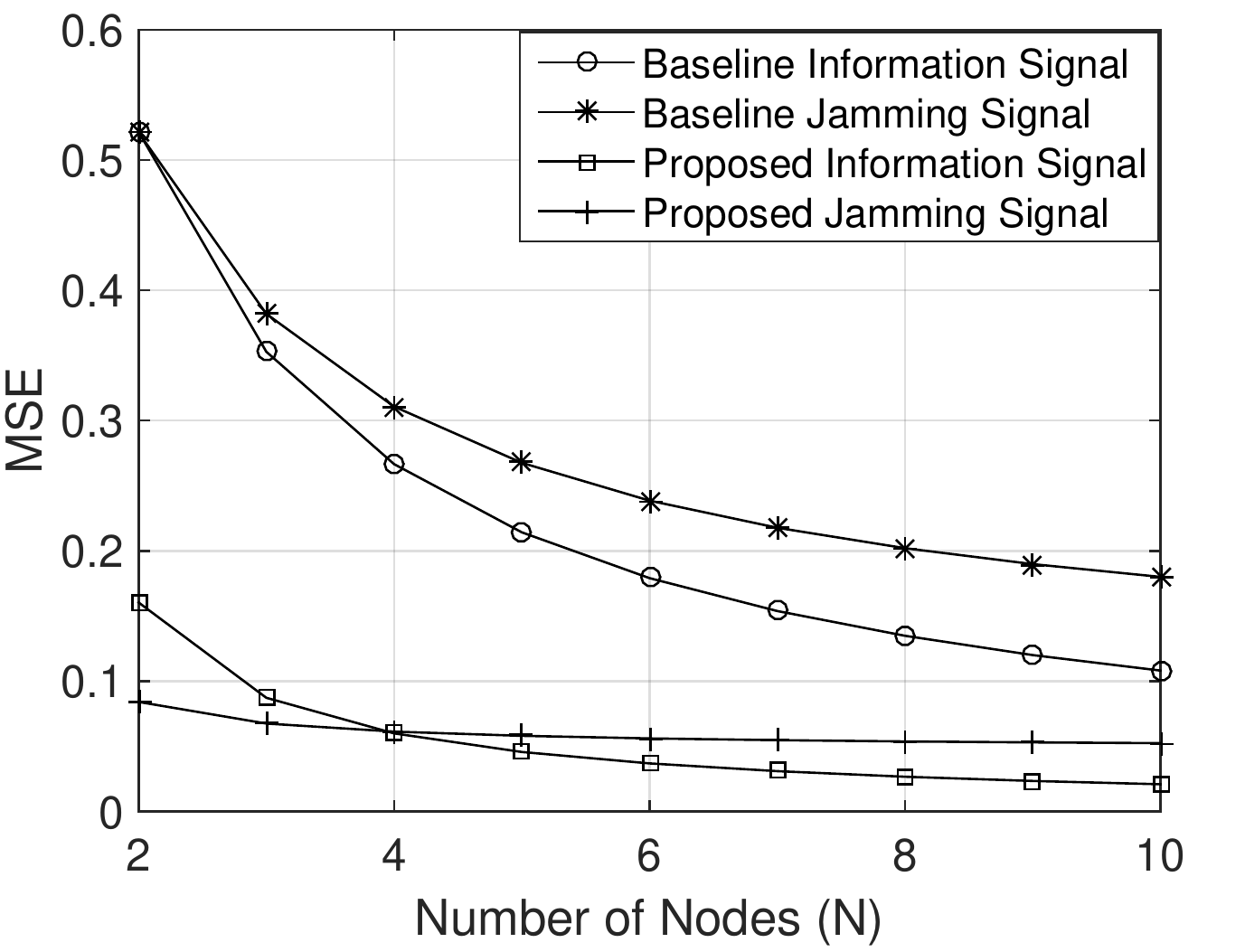}
\caption{Results for the AWGN channel.}\label{fig:MSE_AWGN}
\end{figure}
\begin{figure}[t]
	\includegraphics[width=0.99\linewidth]{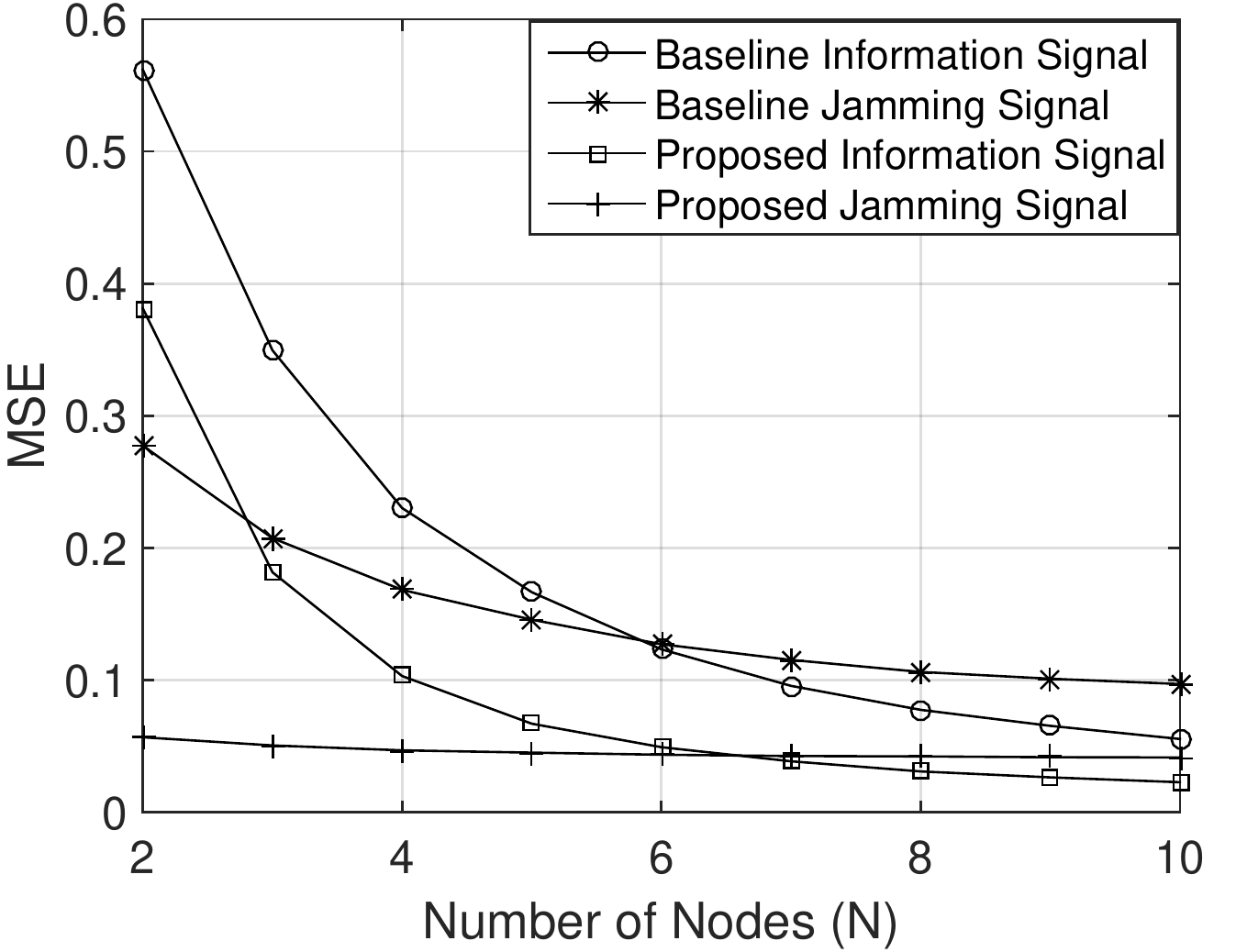}
	\caption{Results for the Rayleigh fading channel.}\label{fig:MSE_RAYLEIGH}
\end{figure}
\begin{figure}[t]
	\includegraphics[width=0.99\linewidth]{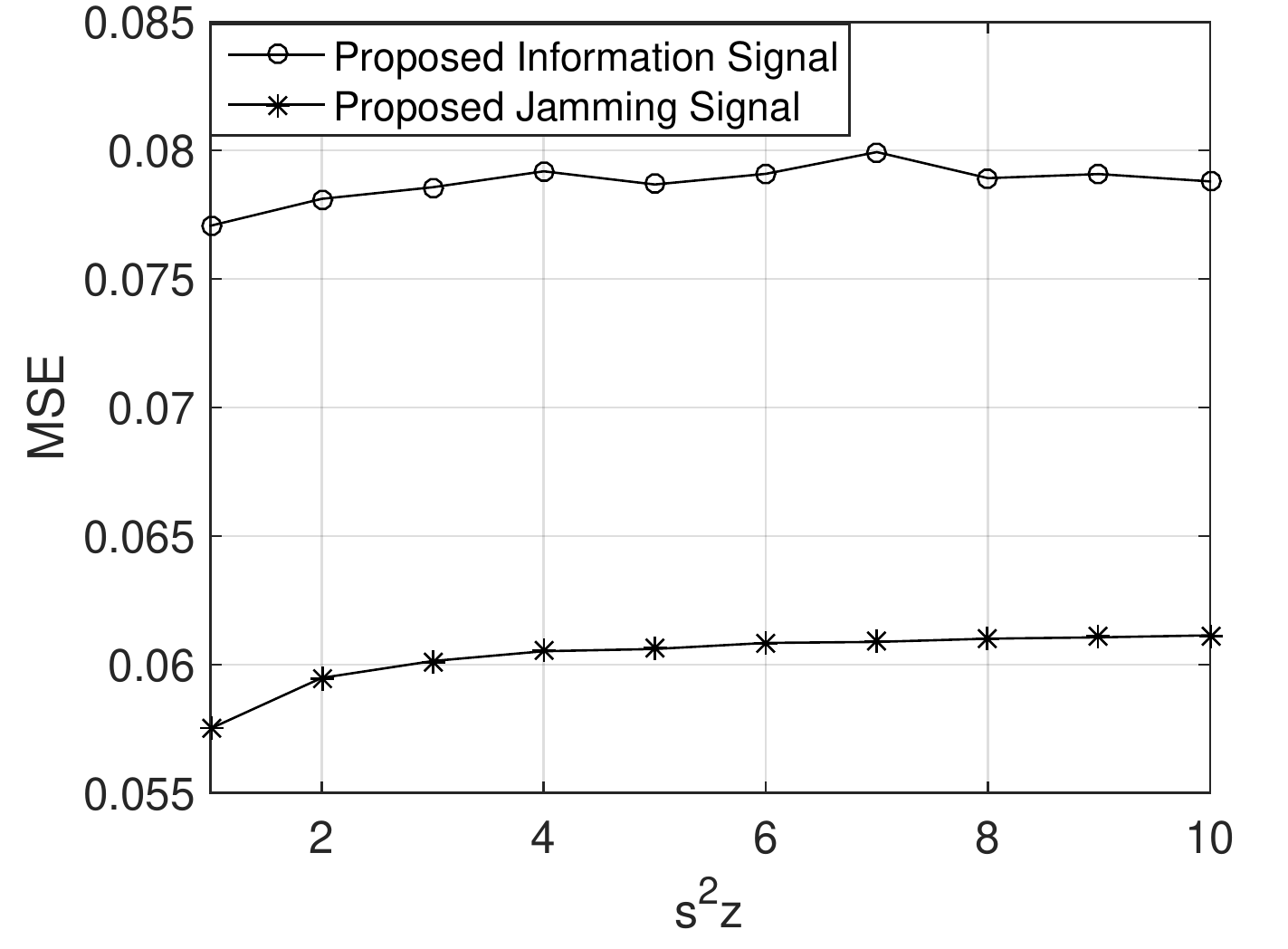}
	\caption{Power of the jamming signal increases in the proposed system for N=5.}\label{fig:MSE_vs_sigmaz_RAYLEIGH}
\end{figure}
\begin{figure}[t]
	\includegraphics[width=0.99\linewidth]{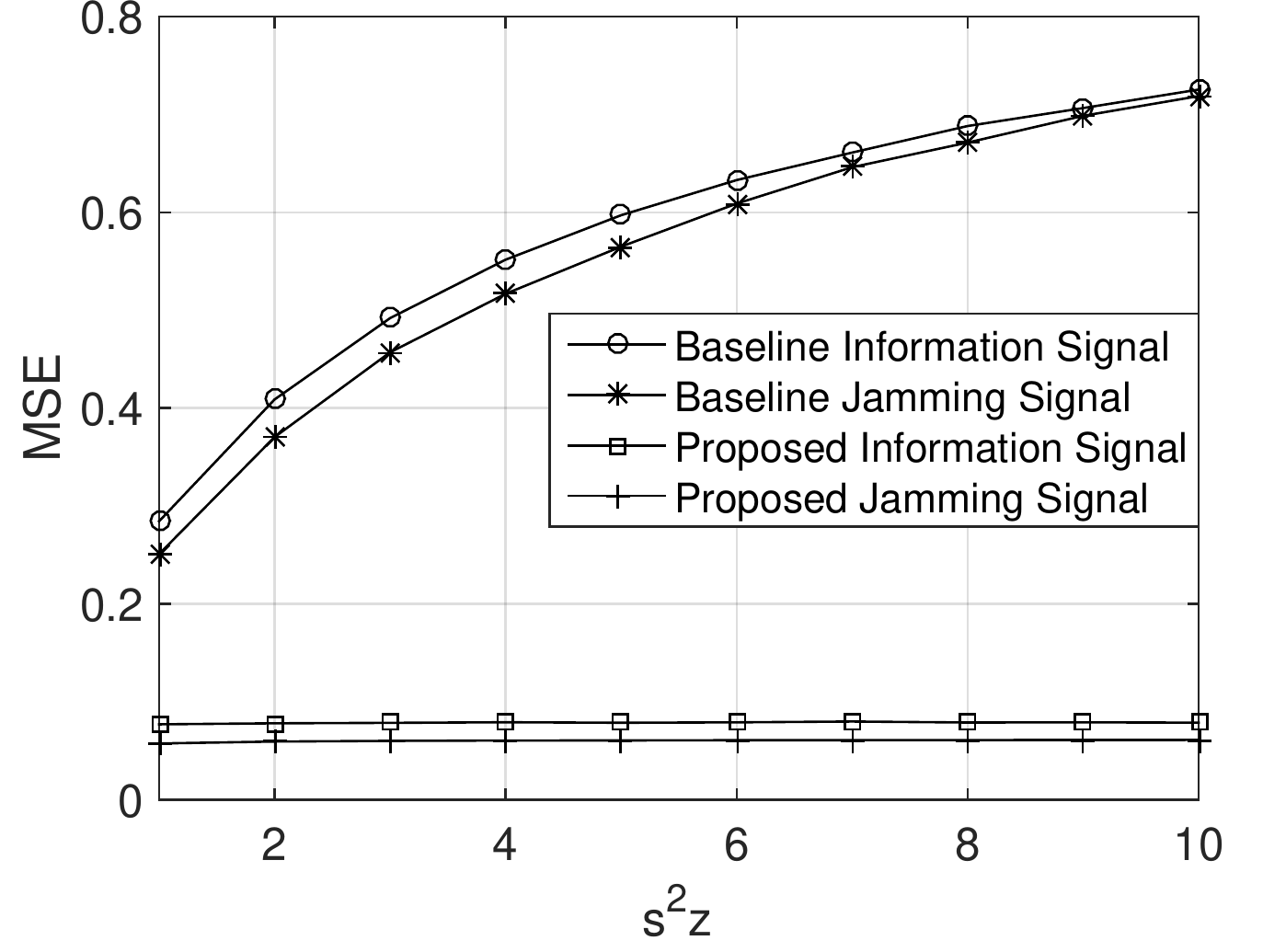}
	\caption{Power of the jamming signal increases in both systems for N=5.}
	\label{fig:MSE_vs_sigmaz}
\end{figure}

\subsection{Results for an AWGN channel} In Fig.~\ref{fig:MSE_AWGN} we present the results for the $MSE_u$ and $MSE_{zi}$ for the proposed and baseline systems. We observe that in the baseline system the $MSE_u$ and the $MSE_{zi}$ for $N=2$ nodes start at the same value. This is what we expect to observe because only \eqref{eqn:2} is available for $u$ and $z_i$ (and $h_i$=1). As we add nodes the two MSE's improve and the MSE of the information $u$ enjoys higher improvements with every new node. For the proposed system our results are much better as we have also the observations from the first time slot for every node and we can estimate and isolate better the jamming signal that eventually results in a better estimation of the information $u$. Although we have better MSE's for both estimated parameters we observe a behavior that requires some further explanation. As we observe in Fig.~\ref{fig:MSE_AWGN} for the proposed system for a number nodes $N=2,3$, the $MSE_{zi}$ is better than the $MSE_u$. This indicates that one can estimate better the different jamming signals for every node than the common information $u$ for all nodes but this is not the case. The reason for this behavior is that the information signal that we are trying to estimate is common for all nodes but the jamming signal $z_i$ is different for every node and contains the unknown channel $h_{ji}$ and the real jamming term $z$. So it is easier for us to estimate a range of values $z_i$ than a discrete value $u$. 
	
\subsection{Results for Rayleigh fading channel}
When the channels between the master node and the other nodes are Rayleigh fading $h_i$ takes random values. We adopt the same assumptions for the variance of the information signal and the noise. In Fig.~\ref{fig:MSE_RAYLEIGH} we present the results for $MSE_u$ and $MSE_{zi}$. We observe that in the baseline system the MSE is greater than the proposed system because in the baseline system we have only the observations of the second time slot for every node so we do not have the ability to estimate the jamming signal. In both systems the $MSE_u$ that is achieved for $N \ge 4$ is adequate for a communication system. The final thing that we observe is that for a small number of nodes the estimation of the jamming signal seems to be better than that of the information. The information signal $u$ that we want to estimate is common for all nodes but the jamming signal is just a different term $z_i$ which contains also the unknown channel $h_{j,i}$ for every node. That means that with the same two observations for every node we are estimating from a set of two possible discrete BPSK values for $u$ (effectively detecting the signal), and simultaneously we estimate $z_i=h_{j,i}z$ (and not $z$ which might also be a discrete modulated signal). The $MSE_{zi}$ has low values even for small $N$. As the number of nodes increases the observations from the different nodes for the information signal $u$ increase leading to a $MSE_u$ that is lower than $MSE_{zi}$. This is achieved for $N \ge 6$.

\subsection{Results for MSE vs $\sigma_{z}^2$}
In our next set of results we assume a constant number of nodes $N=5$ and we vary $\sigma_{z}^2$ between $1$ to $10$. In Fig.~\ref{fig:MSE_vs_sigmaz_RAYLEIGH} we observe that in the proposed system that we have two observations for every node, as $\sigma_{z}^2$ increases,  both $MSE_u$, $MSE_{zi}$ remain practically in the same low desirable value below $0.1$. That means that our system is not vulnerable to jamming, and as the power of the jamming signal $\sigma_{z}^2$ increases the system responds and estimates the information signal $u$ in a very efficient way. In Fig.~\ref{fig:MSE_vs_sigmaz} we observe the difference between the baseline and the proposed system. Here as $\sigma_{z}^2$ increases (power of jamming increases) we observe a massive increase in $MSE_u$ and $MSE_{zi}$. These results illustrate the importance of the observations in~\eqref{eqn:1} for every node. In the baseline system that we practically cannot use these observations we have only \eqref{eqn:2} for every node. That means that we have no more information for every $z_i$ and when this jamming signal has higher power than the information signal we cannot isolate and estimate the later.

\section{Simulation Results for Vehicular Channel}
\label{sec:simulation}
In this section we seek to evaluate the performance of the speed estimation algorithm in conjunction with the jamming signal estimation algorithm. For this purpose, we used the Simulation of Urban Mobility (SUMO) tool and OMNET++/VEINS ~\cite{veins}. SUMO is adopted as our traffic simulator and OMNET++ is used to simulate wireless communication. Furthermore, the GEMV
(a geometry-based efficient propagation model for V2V) \cite{GEMV-BOban} tool was integrated into the VEINS network simulator for a more realistic simulation of the PHY layer~\cite{mimo-kosmanos}. For describing the modeled area GEMV uses the outlines of vehicles, buildings and foliage. Based on the outlines of the objects, it forms  R-trees. R-tree is a tree data structure in which objects in the field are bound by rectangles and are hierarchically structured based on their location in space. Hence, GEMV employs a simple geometry-based small-scale signal variation model and calculates the additional stochastic signal variation and the number of diffracted and reflected rays based on the information about the surrounding objects. 

\subsection{Cooperative Jammer Speed Estimation Results}
\begin{figure}[t]
\includegraphics[width=0.99\linewidth]{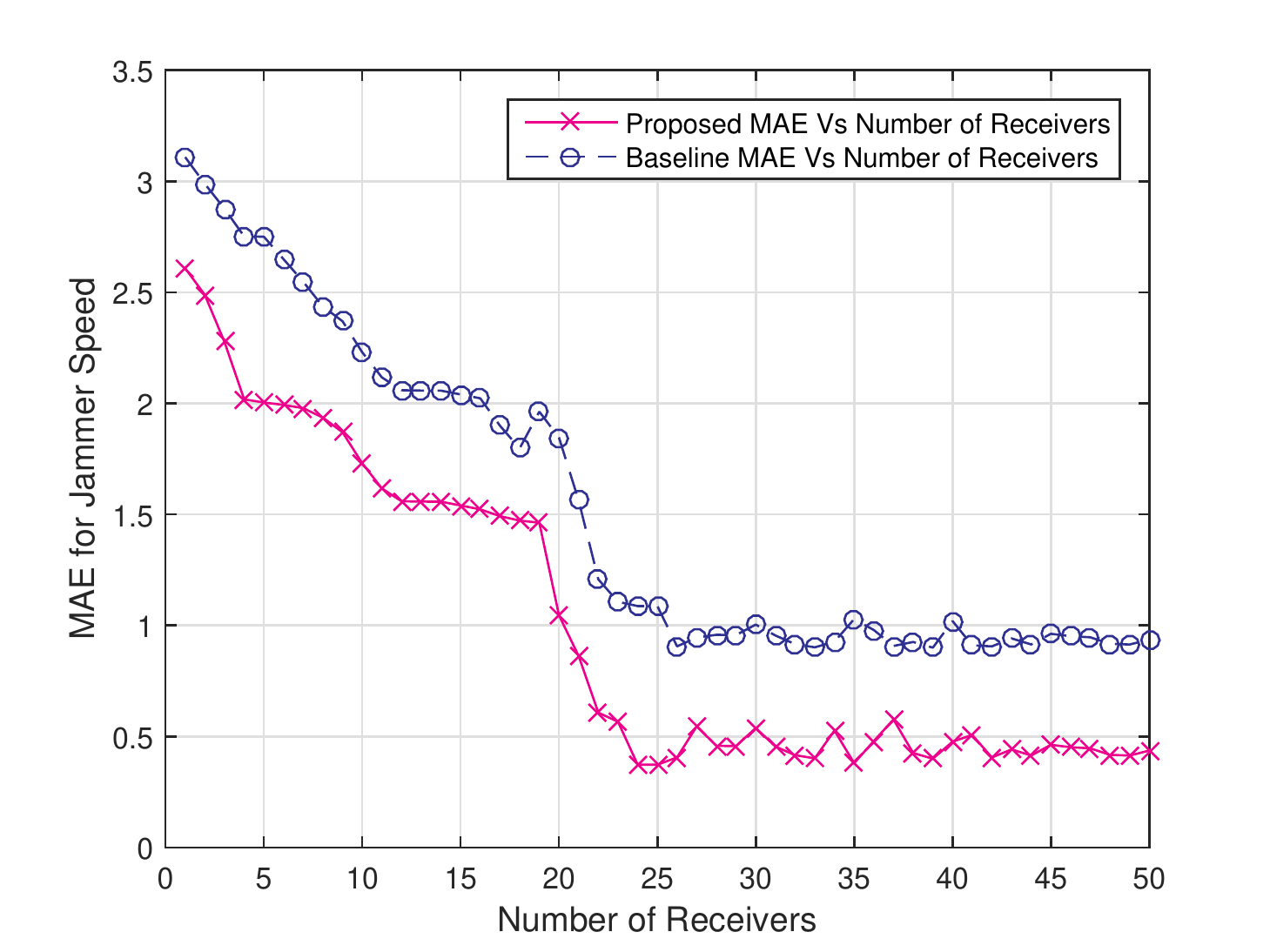}
\caption{Comparison between the proposed and the baseline method for the MAE of the jammer speed estimation using a different number of receivers.}
\label{fig-jammer-speed}
\end{figure}

We present our results in terms of the Mean Absolute Error (MAE) between the real value of the speed $u_{j}$  and the estimated mean of the jammer speed in~\eqref{sum1},~\eqref{sum2} using the estimated results from multiple receivers $\hat{u}_{j}$:
\begin{equation}
MAE=|\hat{\bar{u}}_{j}-u_{j}|
\end{equation}
The MAE is calculated for both baseline and the proposed two-stage transmission scheme. These estimated MAE values are presented in Fig.~\ref{fig-jammer-speed} for both baseline and the proposed two-stage transmission scheme with a different number of receivers in the interval [0,50].  In this experiment we also assume that the jammer approaches the platoon of vehicles with a maximum speed of $28$ km/h and the receivers move with random speeds that belong in the interval [30,38] km/h. By setting the number of receivers to $N=50$ nodes, we observed that the jammer can effectively communicate with only $25$ out of $50$ nodes based on the GEMV simulator \cite{GEMV-BOban}. Because of its superior performance we use the proposed two-stage transmission scheme for the rest of the experiments.

\begin{figure}[t]
\includegraphics[width=0.99\linewidth]{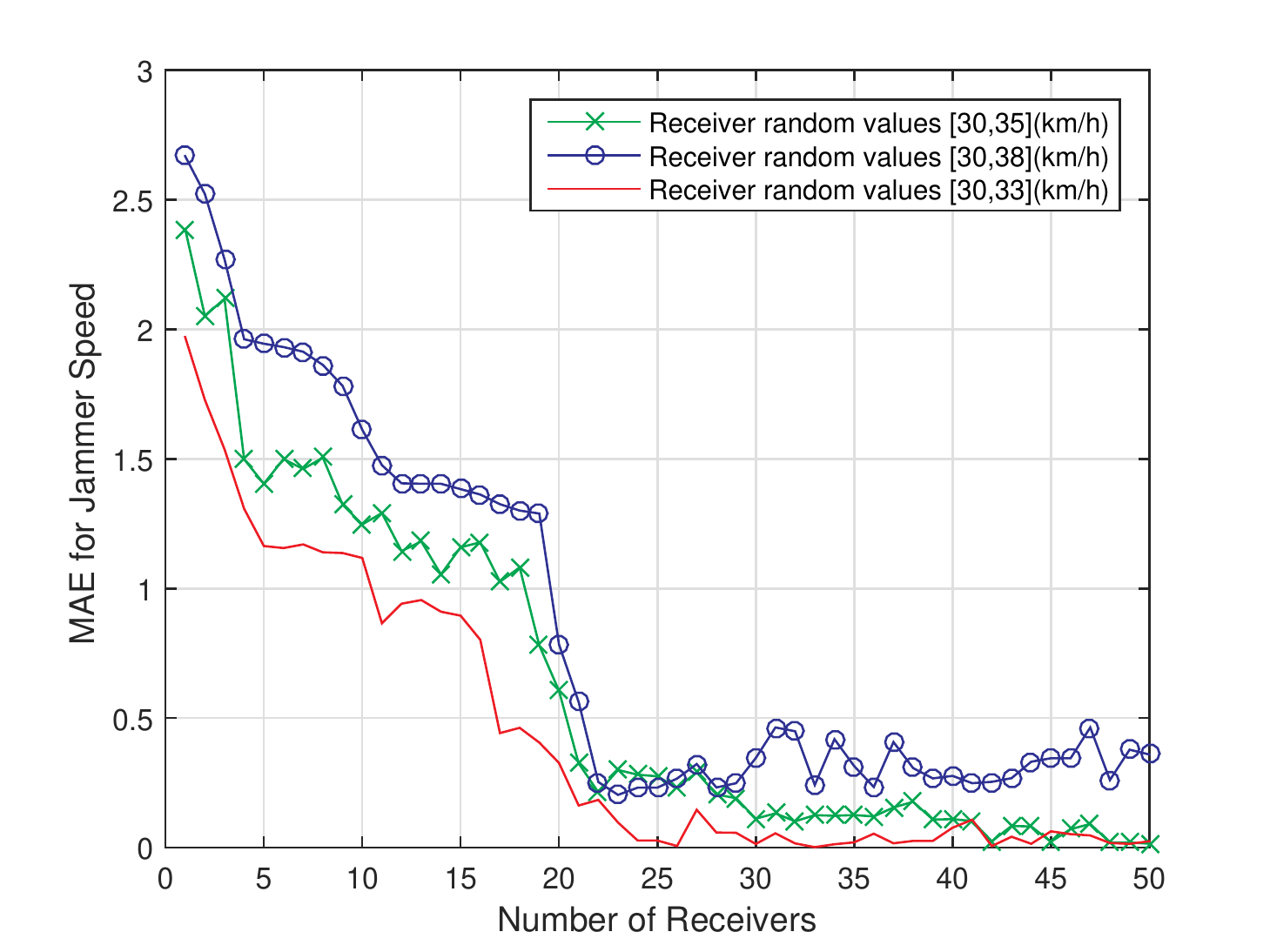}
\caption{MAE of the proposed jammer speed estimation method using different ranges for the speed of the receivers.}
\label{fig-receiver-speed}
\end{figure}

It can be seen in Fig.~\ref{fig-receiver-speed} that after a number of 25 receivers the MAE of the estimated jammer speed converges to a stable value for both systems under comparison. This is because beyond 25, there are no other effective communication pairs between the jammer and the receivers.

But in reality, the members of a platoon of  vehicles never moves at exactly the same speed. So, when we change slightly the range where the speed of the receivers can vary, we observe in Fig.~\ref{fig-receiver-speed} that as this range is narrower, the MAE decreases. This is because in this case the speed deviation of all the receivers will be present as an additional condition in~\eqref{sum1},~\eqref{sum2}. This result is very encouraging since it states that when it is used in vehicle formations that all have the same approximate speed (e.g. platoons), the speed of the jammer can be estimated with improved accuracy.

\begin{figure}[t]
\includegraphics[width=0.99\linewidth]{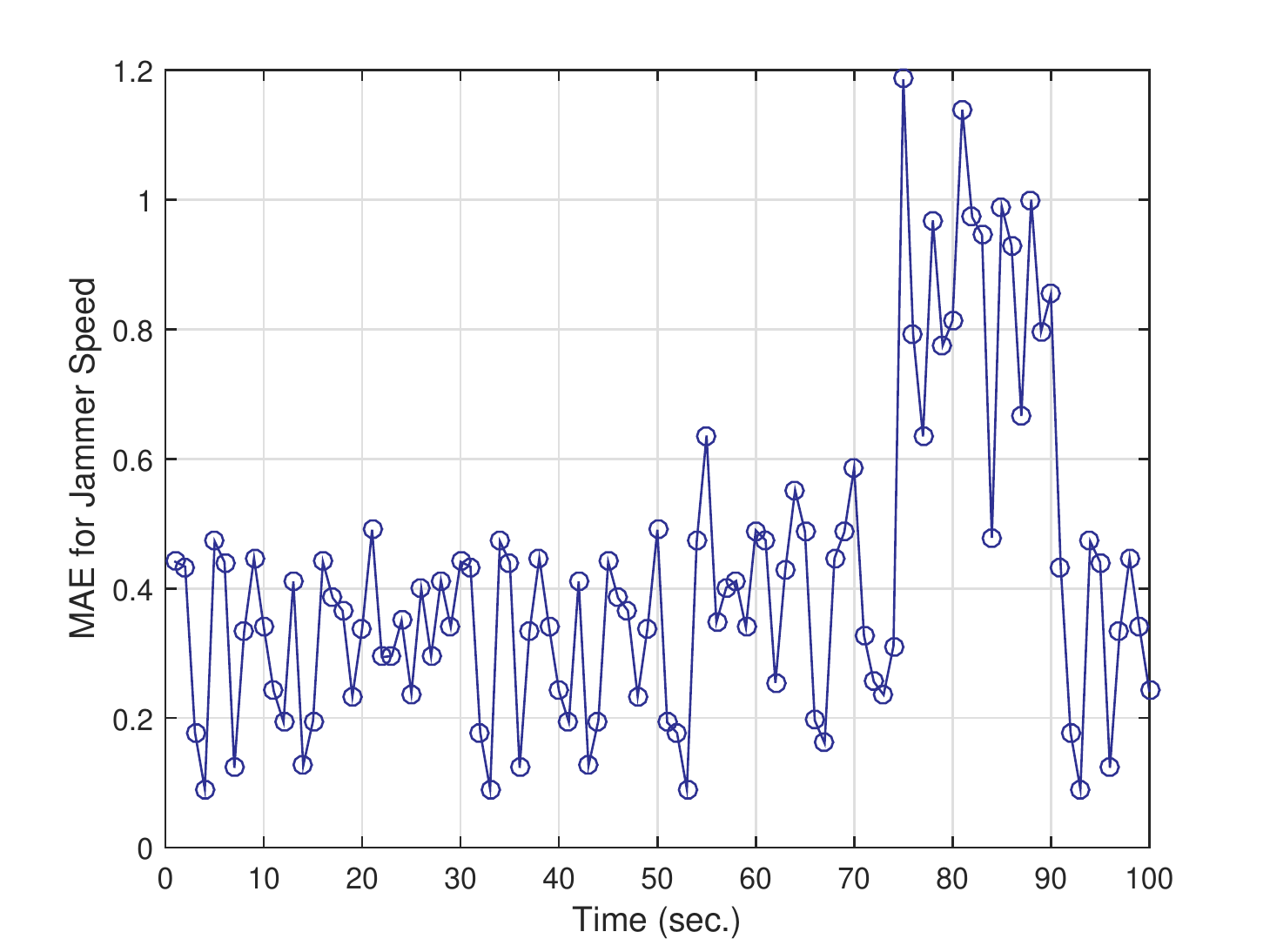}
\caption{Jammer speed estimation for $100$ sec. using an update interval $\Delta{t}=20$ sec. for jamming signal estimation. }
\label{speed-in-time}
\end{figure}

For our next experiment we check the robustness of the proposed distributed system with 20 multiple receivers but over time. We update the jamming signal $z_i$ estimate using~\eqref{eq:estimator1} every $\Delta{t}=20$ sec., while in the intermediate time instants we use the last estimated jamming signal as our current estimate of the jammer speed. The jammer speed estimate that takes place in the time interval $[1,100]$ sec. is presented in Fig.~\ref{speed-in-time}. During $[1,75]$ sec. the jammer moves with a speed of $25$ km/h, while from time $75$ sec. onwards a sharp increase in the speed of the jammer to $50$ km/h is observed. Therefore, for the specific time interval $[75,90]$ sec. the MAE value increases significantly. This is happening because of the jammer speed in the subsequent time instants between $75$ and $90$ sec. is actually the old estimate made at $70$ seconds. This is clearly an approach that may create a stale value for the estimated speed when we have changes.

\begin{figure}[t]
\includegraphics[width=0.99\linewidth]{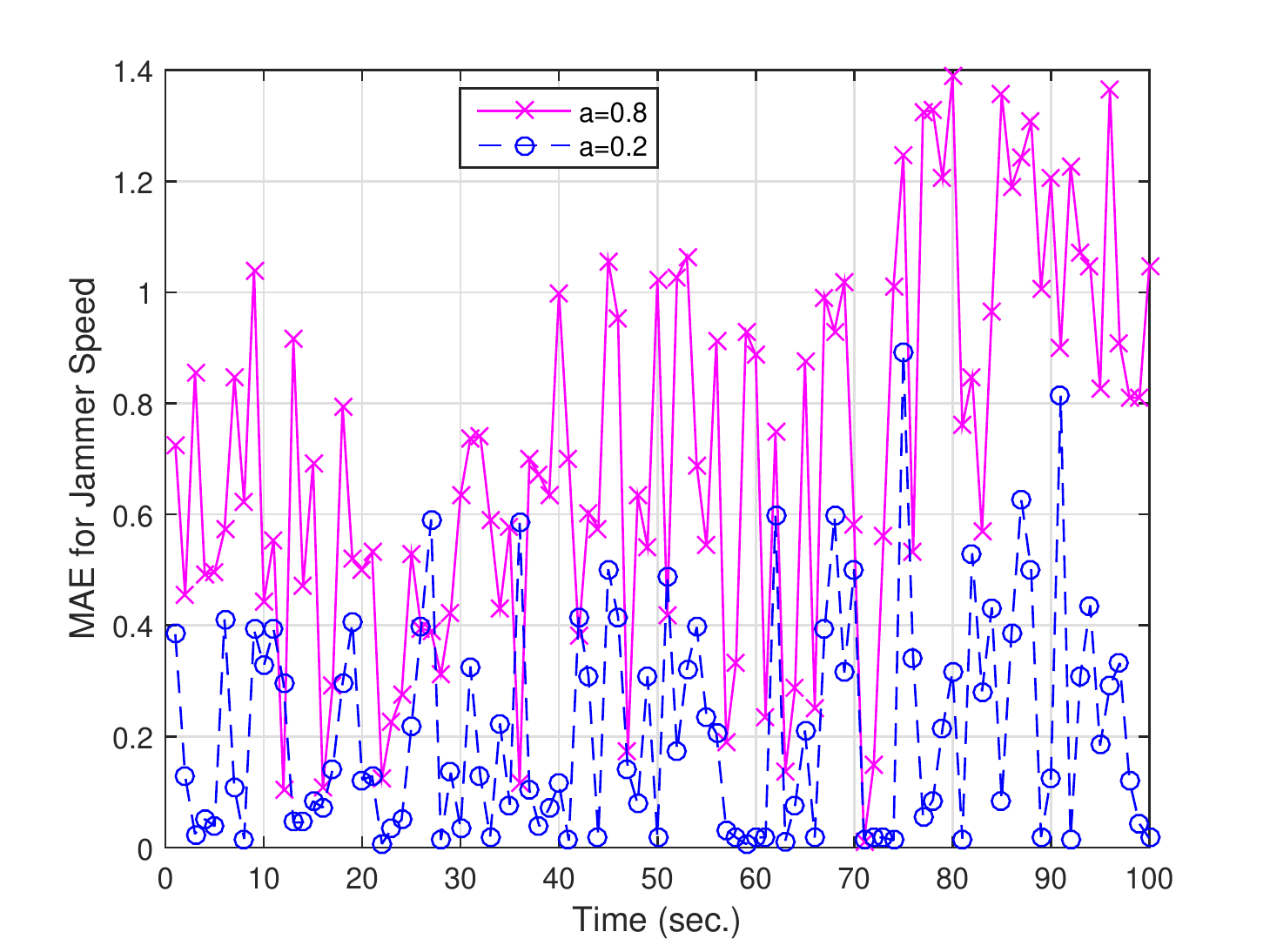}
\caption{Jammer speed estimation for a period of $100$ sec. using different values for parameter $a$.}
\label{speed-a-values}
\end{figure}

To solve the previous problem we apply a smoothing filter for combining the jammer speed estimates across time. In particular we combine the last estimate with the estimated jammer speed at the present time instant $\hat{\bar{u}}_{j}(t)$ as follows:
\begin{equation}
\label{filter}
\hat{\bar{u}}_{j(filtered)}(t)= (1-a) \hat{\bar{u}}_{j}(t) + a * \hat{\bar{u}}_{j(filtered)}(t-1)
\end{equation}
We explored for two extreme values $a=[0.8, 0.2]$ and the results can be seen in Fig.~\ref{speed-a-values}. Observing these results, it is obvious that giving parameter $a$ large values such as 0.8, results in a sharp changes in filtered speed estimate over the entire duration of this specific experiment and especially in the specific time interval [75,90] sec. This is because the speed of the jammer is mainly estimated using the last estimate of its speed. This results in the smoothing filter being unable to estimate the actual instantaneous changes in the speed of the jammer. On the contrary, by giving lower values to parameter $a$ around $0.3$ or $0.2$, all abrupt changes are absorbed by the smoothing filter and so the MAE does not vary significantly over time.

\section{Conclusions}
\label{sec:conclusions}
In this paper, we considered a network when a swarm of nodes that receive an information signal from a master node and a signal from an RF jammer. We proposed first a transmission scheme where the master node remains silent for a slot, and second a joint data and jamming signal estimation algorithm using LMMSE estimation. We derived analytical closed-form expressions for the MSE of our system. Our results indicate that as the number of nodes in the swarm increases, the estimation of both the jamming and information signals is improved significantly. Our results also showed that our proposed transmission scheme is robust against RF jamming attacks since, although the power of the jamming signal ($\sigma_{z}^2$) increases, the $MSE_u$ and $MSE_{zi}$ remains constant. Finally, we proposed a method for combining the jamming signal estimates from the multiple receivers so as to improve the accuracy of the jammer speed estimate. To the best of our knowledge, our proposed scheme is the first distributed estimation scheme for the speed of an RF jammer. The experimental results prove that the speed estimate of the jammer is improved by increasing the number of receivers and the proposed method is particularly suitable for a platoon of vehicles since they use approximately the same speed.


\bibliographystyle{IEEEtran}
\bibliography{paper}	
\end{document}